\documentclass[final,12pt]{elsarticle}

\makeatletter
\def\ps@pprintTitle{%
 \let\@oddhead\@empty
 \let\@evenhead\@empty
 \def\@oddfoot{\centerline{\texttt{zaratan.world}}}%
 \let\@evenfoot\@oddfoot}
\makeatother

\usepackage{graphicx}
\usepackage{amssymb}
\usepackage{amsmath}
\usepackage{listings}
\usepackage{url}
\usepackage{algorithm}
\usepackage{algpseudocode}
\usepackage{subcaption}

\makeatletter
\providecommand{\leftsquigarrow}{%
  \mathrel{\mathpalette\reflect@squig\relax}%
}
\newcommand{\reflect@squig}[2]{%
  \reflectbox{$\m@th#1\rightsquigarrow$}%
}
\makeatother

\lstdefinestyle{solstyle}{
    basicstyle=\ttfamily\small,
    frame=\lines
}

\usepackage{lineno}

\biboptions{round}
\newcommand{\citex}[1]{\citeauthor{#1} \citeyearpar{#1}} 
\newcommand{\citey}[1]{\citeauthor{#1}, \citeyear{#1}} 

\begin{document}

\begin{frontmatter}


\title{Cybernetic Governance in a Coliving House}



\author[1]{Daniel Kronovet}
\author[2]{Seth Frey}
\author[3]{Joseph DeSimone}

\address[1]{Zaratan Coliving - krono@zaratan.world}
\address[2]{UC Davis Department of Communication - sethfrey@ucdavis.edu}
\address[3]{Asmodee Group - j.desimone@asmodee.com}


\begin{abstract}
We report an 18-month field experiment in \textit{distributed digital institutions}: a nine-bedroom Los Angeles coliving house that runs without managers, while sustaining 98\% occupancy and below-market rents.

Drawing on Elinor Ostrom’s commons theory, we outline design principles and three digital mechanisms that form the institutional core: 1) A continuous-auction chore scheduler turns regenerative labor into a time-indexed points market; residents meet a 100-point monthly obligation by claiming tasks whose value rises linearly with neglect. 2) A pairwise-preference layer lets participants asynchronously reprioritize tasks, translating meta-governance into low-cognition spot inputs. 3) A symbolic “hearts” ledger tracks norm compliance through automated enforcement, lightweight challenges, and peer-awarded karma. Together, these mechanisms operationalize cybernetic principles—human sensing, machine bookkeeping, real-time feedback—while minimizing dependence on privileged roles.

Our exploratory data (567 chore claims, 255 heart events, and 551 group purchases) show that such tooling can sustain reliable commons governance without continuous leadership, offering a transferable design palette for online communities, coliving houses, and other digitally mediated collectives.
\end{abstract}

\begin{keyword}
Economics \sep Cybernetics \sep Commons \sep Institutions, Housing
\end{keyword}

\end{frontmatter}

\pagebreak



\section{Introduction}
\label{intro}

There is an aphorism attributed to Dorothy Day of the Catholic Workers Movement: ``Everybody wants a revolution, but nobody wants to do the dishes.'' All too often coliving communities form full of enthusiasm, only to crumble when confronted with the boring but essential regenerative labor of daily life. Individuals shirk responsibilities, leaders burn out, and people leave feeling cynical and defeated. Can we do better?

\vspace{2mm}

This paper will present a case study of Sage House, a non-hierarchical coliving house which self-regulates using computational tools. Sage is a 9-bedroom house in Highland Park, Los Angeles, continuously operating since September 2022. Sixteen people have lived in the house since opening, with an average tenure of 11 months and an overall occupancy rate of 98\%. Sage is ``naturally affordable'' in that it offers low-market rents without reliance on external support -- achieved largely through the operational innovations presented here.\footnote{While not the focus of this paper, our discussion has significant implications for housing affordability. High-touch management can add 10\% or more to housing costs.} The house is co-owned by one of the authors, Daniel Kronovet, who also led the renovation, lived in the house in its first year of operation, and continues to play a limited administrative role. Much of the ethnographic perspective on Sage is drawn from Daniel’s experience.

The computational suite, known as Chore Wheel (\citey{kronovet:2020}), was designed in close reference to Elinor Ostrom’s theory of institutional design. With the aid of this suite, the regenerative labor burden is fairly and transparently distributed among the residents, with minimal need for unsustainable individual leadership or intensive deliberative processes. This low contribution requirement, approximately 2-4 hours of chores and one 90-minute meeting per month, has made Sage resilient to turnover and interpersonal conflict and allows the house to operate continuously at low cost.

To frame our case study, we will explore \textbf{two motivating themes}: that of the institutional role of leadership, and of the regulatory potential of cybernetic systems driven endogenously by time.

\vspace{2mm}

\textbf{Regarding leadership}, \citex{ostrom:1990} discusses the theoretical challenge associated with institutional production. Defining institutions as “prescriptions [used] to organize ... repetitive and structured interactions," Ostrom reasons that while agents can be incentivized \textit{within} institutions, they cannot be incentivized to \textit{produce} them. Institutional production, Ostrom concludes, comes from hard-to-theorize personal motivation, such as the drive for status or for creative expression. Another term for institutional production, for going beyond incentives, is leadership.

In the context of this case study, we will characterize \textit{leadership} as the performance and communication of high-order cognition (multi-step reasoning) on behalf of a group. In contrast, \textit{non-leadership} is characterized as low-order cognition (one- or two-step reasoning) in response to environmental stimuli, including communications by leadership. Leadership is framed as an act of \textit{institutional production} enabling non-leaders to perform coordinated activities through low-order responses to evolving informational contexts.

\vspace{2mm} 

Approaching the question of leadership from the perspective of game design, \citex{koster:2018} describes how certain game mechanics allow for low-trust ``parallel play'' (e.g. soccer) while others \textit{require} high-trust coordinated action (e.g. trapeze). One of Koster’s conclusions is that parallel-play designs do not \textit{preclude} the emergence of high-trust coordinated action, only that such games can be enjoyed without it. Compared to high-trust games, parallel-play games are more accessible, and thus more robust as institutions, without being limited in their potential for enabling deep experience.

We suggest that over-reliance on leadership represents a type of institutional fragility. Good leaders are rare, and communities which rely on consistent leadership are vulnerable to disruption. One can pay for leadership, but it is expensive, and financial incentives can introduce alignment problems. Further, explicit assignment of leadership implicitly assigns others to non-leadership. The more that \textit{all} participants can \textit{temporarily} assume leadership roles, the greater the total leadership resources the organization can draw upon (\citey{laloux:2014}). A resilient institution is one which can \textit{benefit} from periodic expressions of leadership without being \textit{dependent} on them.

\vspace{2mm}

\textbf{Regarding cybernetics}, Beer (1973) advocates for computers as \textit{essential} tools in a technologically advanced society. For Beer, the computer is a machine which can make the static \textit{dynamic}, and thus more concisely representative of an evolving world. By making complex information legible, and by incorporating ongoing human feedback, the computer takes on a meaningful coordinating role.

In his work, Beer draws on the influence of pioneering cyberneticist W. Ross Ashby, who as \citex{lewis:2024} noted, exerted significant influence on the Ostroms themselves. Ashby is known for his ``Law of Requisite Variety,'' which states that the number of possible states (variety) in the system must be matched by the number of possible states in the regulator. Succintly, \textit{a regulator must be as expressive as that which it regulates}. For Beer, the use of computers as dynamic tools allows them to more fully capture the \textit{variety} inherent in a complex society, turning them into indispensable aids for navigating an ever-changing world.

In conventional organizations, resources are \textit{at rest} until energy is spent putting them into motion: we ``go to mangement'' to beg funding for our projects. In contrast, through the use of computers, resources can be kept in \textit{perpetual motion over time}, enabling new and more expressive classes of asynchronous interactions (\citey{rea:2018}, \citey{tan:2023}). As Beer writes, the static quarterly report can be replaced by real-time representations of value flows, enabling organizational members to respond proactively and autonomously to new information and opportunities.

\vspace{2mm}

\textbf{At the intersection} of these thematic lines is the idea that computational systems can be developed which \textit{preserve and integrate continual but intermittent contributions of leadership}, while \textit{providing useful structure continuously to non-leaders}. A defining feature of these systems is the use of time and feedback loops to turn the computer into an active and enabling force in organizational life. Through the application of these design techniques, the leadership burden can be both reduced and distributed more evenly among participants, allowing the development of more resilient and reliable institutions.
\section{Design Principles}
\label{principles}

\citex{ostrom:1990} defines a \textit{common-pool resource (CPR)} as one which is both \textit{rivalrous} (two people cannot use it at once) and \textit{non-excludable} (it is difficult to keep people from using it), with forests and fisheries being prototypical examples. In the coliving setting, the resource in question is \textit{cleanliness}, or more generally, \textit{order}. Residents ``consume order'' during the day – whether by using a dish, tracking dirt on the floor, eating food from the fridge, or making noise in common space. Each of these actions constitutes, in ways large or small, a consumption of shared order for personal benefit. \textbf{Sustaining this CPR requires residents to continually produce order such that homeostasis is maintained.}

Seen through this lens, a coliving house becomes a type of commons, and coordinating the ``production of order'' (regenerative labor) becomes the problem to be solved. \citex{ostrom:1990} famously articulated \textbf{eight design principles} seen in effective CPR-managing institutions. We will introduce these principles and summarize how they are realized at Sage House. In section \ref{mechanism}, we will discuss how these principles are implemented by Chore Wheel. These principles are:

\begin{enumerate}
    \item \textbf{Clearly defined group boundaries}
    \begin{itemize}
        \item All participants are house residents legally bound by the lease
        \item All participants are members of a shared communications platform
    \end{itemize}
    
    \item \textbf{Mechanisms adapted to local conditions}
    \begin{itemize}
        \item All mechanisms address specific problems present in coliving
        \item All mechanisms leverage easily-accessible local information
    \end{itemize}
    
    \item \textbf{Members participate in decision-making}
    \begin{itemize}
        \item Resource-spending priorities are determined by residents
        \item Behavioral norm-setting is performed by residents
    \end{itemize}
    
    \item \textbf{Effective monitoring}
    \begin{itemize}
        \item Resource claims are validated by residents using local information
        \item Contribution requirements are enforced automatically
    \end{itemize}
    
    \item \textbf{Graduated sanctions}
    \begin{itemize}
        \item Intermittent norm violations result in symbolic penalties only
        \item Significant or chronic norm violations result in financial penalties
    \end{itemize}
    
    \item \textbf{Cheap and accessible conflict resolution}
    \begin{itemize}
        \item Dishonest resource claims are easily invalidated by other residents
        \item Personal disputes are resolved with a symbolic challenge process
    \end{itemize}
    
    \item \textbf{Self-determination of the community}
    \begin{itemize}
        \item Penalties are backstopped by court-enforceable lease language
    \end{itemize}
    
    \item \textbf{Multiple layers of nested enterprises}
    \begin{itemize}
        \item Not currently applicable
    \end{itemize}
\end{enumerate}

In addition to Ostrom’s eight principles, we introduce a further four design principles for \textbf{\textit{distributed digital institutions}}, described in reference to the IAD framework of \citex{ostrom:2005}:

\begin{enumerate}
    \item \textbf{No managers or privileged roles}
    \begin{itemize}
        \item There should be as few differentiated \textbf{positions} as possible
        \item Participants can flexibly ``parallel play'' leadership roles
    \end{itemize}
    
    \item \textbf{Simple and intuitive inputs}
    \begin{itemize}
        \item \textbf{Choices} made with \textbf{local information} and low-degree cognition
        \item New participants can start engaging quickly, improving resiliency
    \end{itemize}
    
    \item \textbf{Humans for sensing, machines for bookkeeping}
    \begin{itemize}
        \item \textbf{Aggregation} and high-order cognition done computationally
        \item Only humans make subjective judgments, preserving legitimacy\footnote{\citex{zuboff:2018} describes how the uncritical use of surveilled data and  statistical inference (machine learning) in technical systems has resulted in higher levels of coercion and lower levels of human agency. By preferring deterministic processes and limiting the use of probabilistic inference, the technical system remains socially embedded.}
    \end{itemize}
    
    \item \textbf{Continuously available, asynchronous processes}
    \begin{itemize}
        \item \textbf{Choices} are available and \textbf{aggregation} occurs continuously
        \item Adaptive lazy consensus lowers coordination overhead\footnote{``Adaptive lazy consensus'' is an alternative to a quorum, in which a certain number of \textit{positive votes} within a time frame are needed to approve an action, rather than a certain number of votes \textit{in total}. Analogous to an ``activation energy'' in chemistry.} 
    \end{itemize}
\end{enumerate}

By leveraging computation to perform high-order cognition and to construct decision environments which continually engage non-leader participants, the leadership requirement to sustain institutions can be reduced. Systems designed according to these principles can be meaningfully used by a diversity of individuals under real-world conditions, enabling cooperation among large and heterogeneous populations. Further, as we shall see in section \ref{mechanism}, these design principles can be used to ``wrap'' more complex mechanisms, exposing significant regulatory power to participants while abstracting away much of the technical complexity. On balance, we believe these design principles and constraints represent a compelling framework for developing a wide class of human-centered computer applications.
\section{Mechanism Overview}
\label{mechanism}

We now describe three of Chore Wheel’s mechanisms for ``producing order:'' \textit{Chores}, \textit{Hearts}, and \textit{Things}. All three tools were designed in reference to the principles presented in section \ref{principles}.

\begin{figure}[ht]
    \centering
    \begin{subfigure}[b]{0.3\linewidth}
        \centering
        \includegraphics[width=\textwidth]{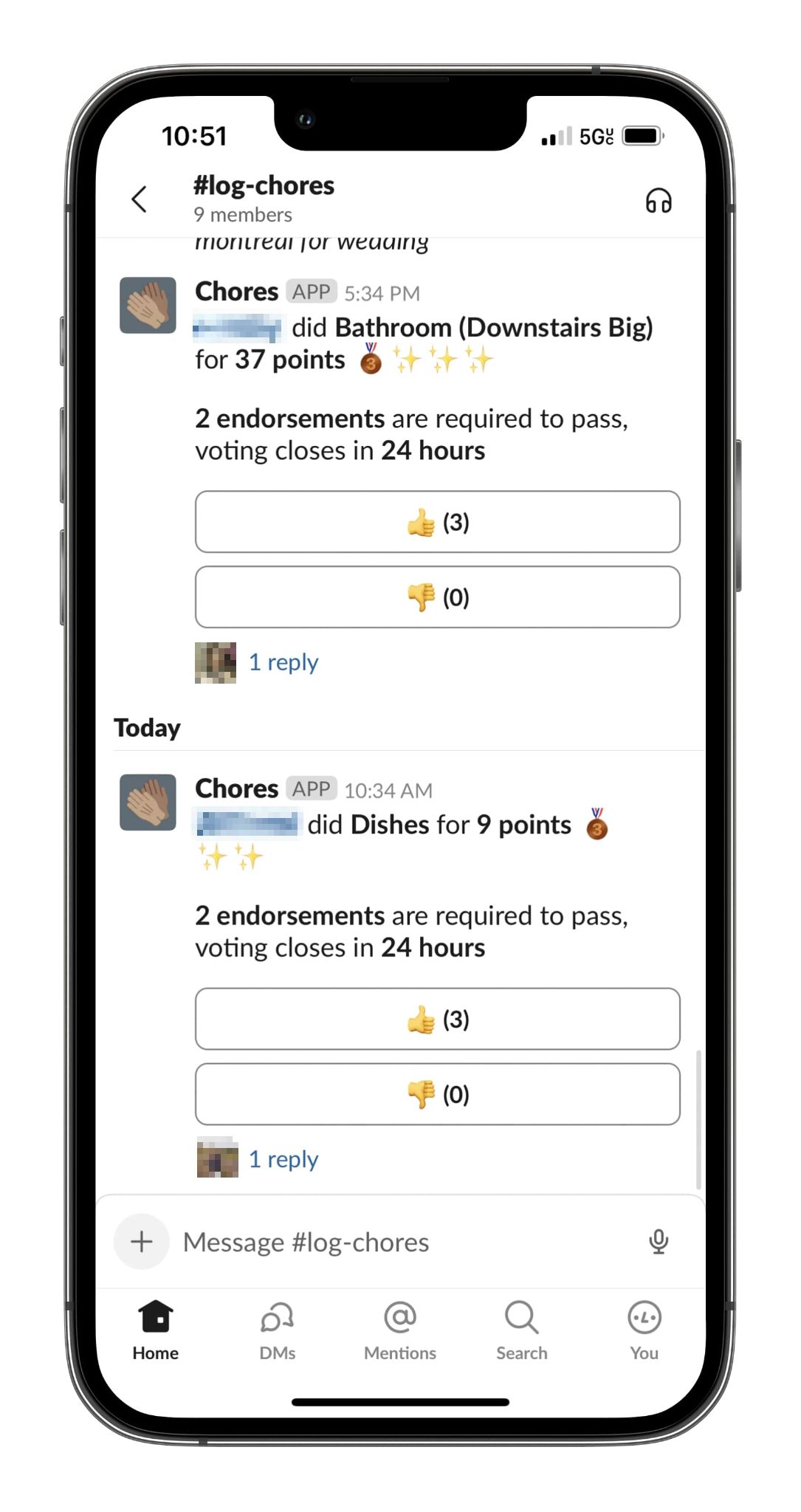}
        \caption{\textit{Chores}}
        \label{fig:chores}
    \end{subfigure}
    \hfill
    \begin{subfigure}[b]{0.3\linewidth}
        \centering
        \includegraphics[width=\textwidth]{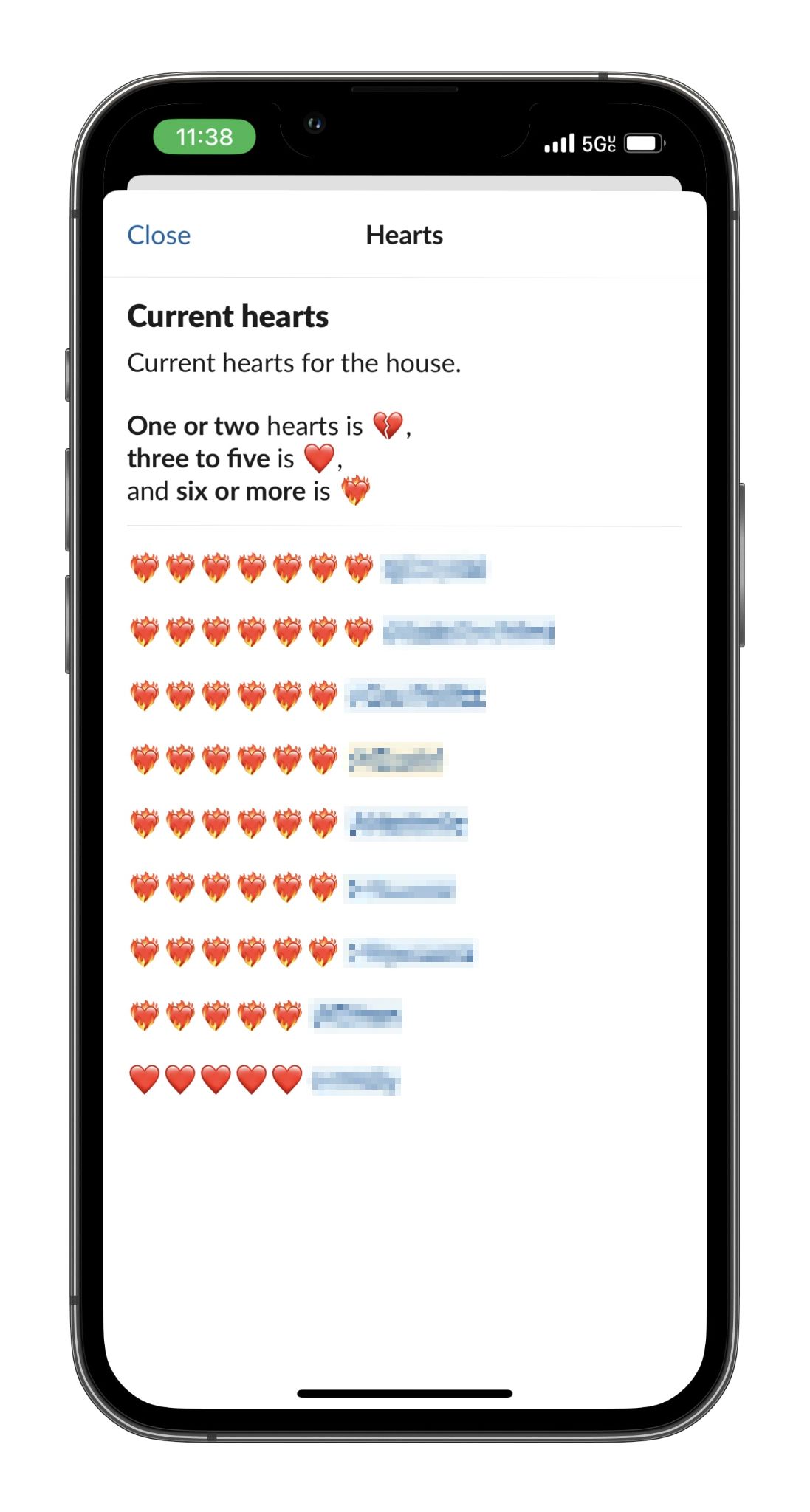}
        \caption{\textit{Hearts}}
        \label{fig:hearts}
    \end{subfigure}
    \hfill
    \begin{subfigure}[b]{0.3\linewidth}
        \centering
        \includegraphics[width=\textwidth]{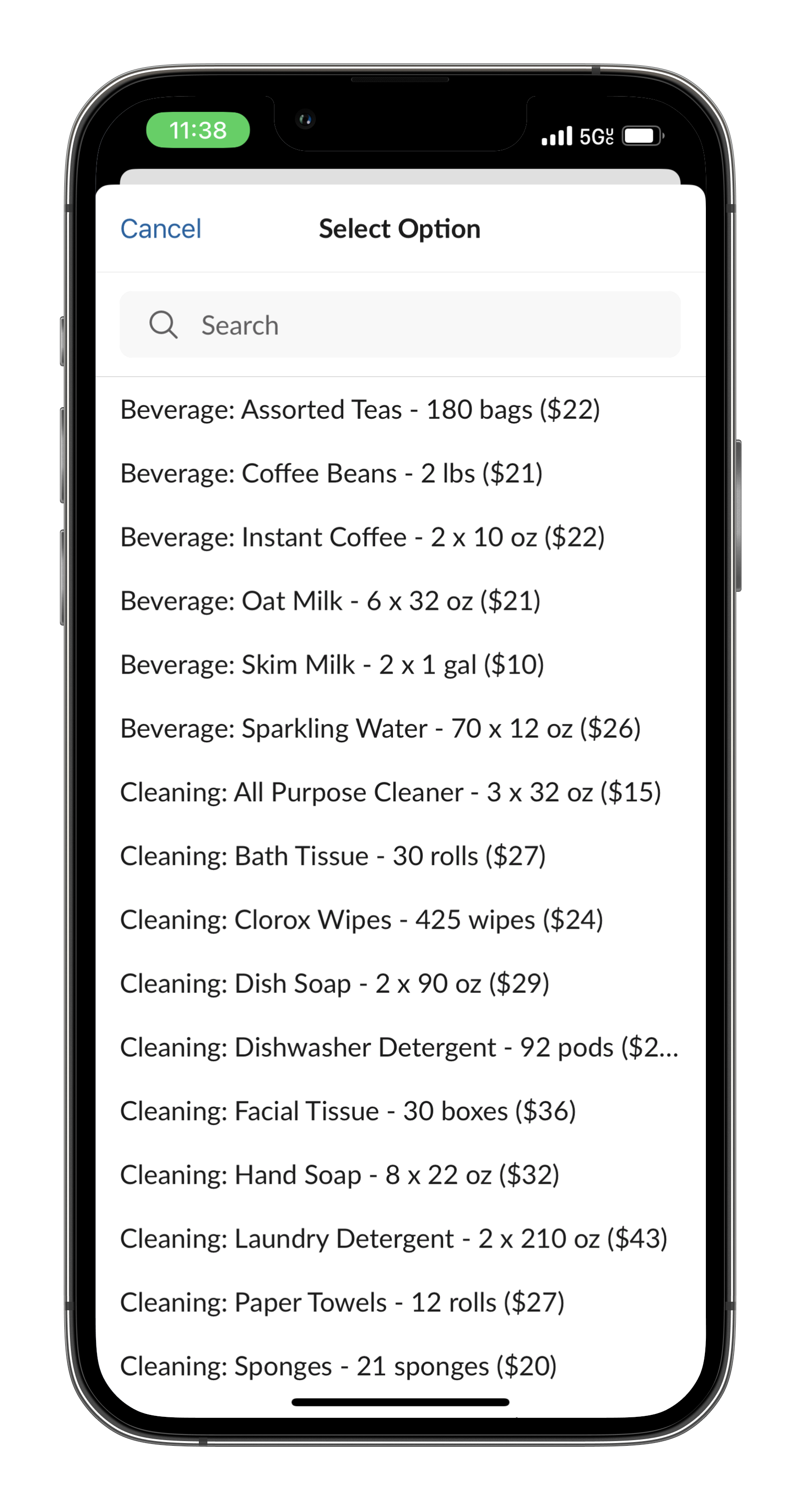}
        \caption{\textit{Things}}
        \label{fig:things}
    \end{subfigure}
    \caption{Chore Wheel interface examples. Figure \ref{fig:chores} shows chore claims being verified in a shared communications channel. Figure \ref{fig:hearts} shows the public ``hearts board'' for all residents. Figure \ref{fig:things} shows an excerpt from the \textit{Things} purchasing flow.}
    \label{fig:chore-wheel-screenshots}
\end{figure}

The mechanisms are implemented as Slack apps and accessed via a shared Slack workspace, through which governance interactions are blended with general inter-resident communication. This strategy, inspired by the chat-bots used to govern Discord gaming servers, has been effective at reducing barriers to engagement. Interface examples are given in \textbf{Figure \ref{fig:chore-wheel-screenshots}}.

\subsection{Chores}
\vspace{2mm}

\textit{Chores} is the cornerstone of Chore Wheel. It combines a \textit{continuous-auction task scheduler} (Ostrom's ``operational layer'') and a \textit{distributed task prioritization mechanism} (her ``policy layer'') to allocate predefined recurring tasks among house residents.

Prior work on chore division shows that when valuations are static and known \textit{a priori}, one can compute assignments that are pareto-optimal and envy-free-up-to-one-item in the one-shot setting (\citey{ebadian:2022}), and at least proportional in the iterated setting (\citey{igarashi:2024}). Our system tackles the more realistic case in which both chore difficulties and resident utilities drift continuously, and where fairness must emerge dynamically rather than through a single offline optimization. Our setup is as follows:

\begin{enumerate}
  \item All residents owe 100 points per month, earned by doing chores.
  \item Points accrue continuously: the longer a chore remains undone, the more points it is worth.
  \item The monthly points budget is fixed at 100 points per resident; residents govern the \textit{relative accrual rates} of individual chores.
\end{enumerate}

\vspace{2mm}

\textbf{The continuous-auction task scheduler}\footnote{We use ``scheduler'' in the computing sense of ``choosing which task to perform next,'' not in the colloquial sense of ``setting a time and date in advance.''} works by continuously increasing the points value of a task until it is performed by a resident who perceives that the reward in points justifies the effort needed to complete the task. The core modeling assumption, made via social-science analogy to the second law of thermodynamics, is that the longer it has been since someone did the dishes, or swept the floor, the more dishes there are in the sink, and the more dirt there is on the floor. As such, a \textit{continuous, linear} increase in points over time effectively models \textit{stochastic, non-linear} increases in mess and disorder in real physical space.

All task claims are presented to the residents for verification in the form of upvotes or downvotes. Due to the closely-shared physical environment, attempts to dishonestly claim a task are easy to identify and reject. As such, the task scheduler achieves high levels of fairness, flexibility, transparency, and accountability, without dependence on ongoing leadership inputs.

\vspace{2mm}

\textbf{The distributed task prioritization mechanism} builds on \citex{kronovet:2017} and \citex{kronovet:2018}, leveraging pairwise preference inputs to produce priority distributions over sets of tasks, governing the rate at which individual chores gain points. As we shall see in section \ref{analysis}, a higher priority can reflect a more complex task, a simple task which must be performed frequently, or both. The core insight motivating this mechanism is that while synchronously producing a full set of priorities across all tasks in the abstract is cognitively challenging, asynchronously producing pairwise judgments in response to local information is cognitively simple. For example: ``I see the dishes are consistently dirty, and have noticed that the yard is fairly clean. I would like to \textit{prioritize} dishes, and \textit{deprioritize} yardwork.'' By replacing a collective decision process of high degree with an individual decision process of low degree, the marginal task prioritization problem is simplified, without sacrificing the end goal of a total prioritization.

Further, the exchange of \textit{static} for \textit{dynamic} prioritization, in the form of ongoing ``spot adjustments,''  allows the system to continuously and incrementally adapt to changing sentiment and circumstance. By way of these just-in-time adjustments, the total prioritization can flexibly adapt to individuals moving in or out or experiencing changes in their personal preferences. Through use of these techniques, participants are \textit{continuously enfranchised}.

\vspace{2mm}

In addition to the two mechanisms discussed above, residents can propose amendments to the chore list itself. This higher-order act is an example of temporarily ``performing leadership'' as once the new chore is established, the leadership task is complete.

Taken together, these constituent mechanisms create a ``cybernetic system'' through which participants gain both high levels of flexibility in meeting their commitments, and high levels of agency in exercising influence over their and others’ contributions towards sustaining a shared physical environment.

\subsection{Hearts}
\vspace{2mm}

\textit{Hearts} is the general-purpose behavioral management mechanism of Chore Wheel. While the \textit{Chores} mechanism is task-specific, reflecting a higher level of structure in the underlying problem, \textit{Hearts} provides a relatively general-purpose mechanism for handling the less-structured problem of maintaining intersubjective social norms.

The core unit for \textit{Hearts} is, appropriately, hearts. Drawing intentionally on the gaming vernacular, a ``heart'' is a symbolic token representing the overall quality of a resident’s conduct. Residents begin with a baseline of five hearts, and gain and lose hearts as the result of various processes. There is no penalty for losing a few hearts, and lost hearts regenerate automatically over time. The pathways for gaining and losing hearts are:

\begin{itemize}
    \item \textbf{Gaining Hearts:}
    \begin{itemize}
        \item Earning karma from other residents \textit{(active)}\footnote{Karma is given by appending \texttt{++} to a resident's name. Each month, the residents with the most karma earn hearts.}
        \item Lost hearts regenerate over time, back to baseline \textit{(passive)}
    \end{itemize}
\end{itemize}

\begin{itemize}
    \item \textbf{Losing Hearts:}
    \begin{itemize}
        \item Losing an explicit challenge from another resident \textit{(active)}\footnote{Challenges are ``symmetric'' in that if a resident makes a challenge and loses, they lose hearts. Winning requires majority support as well as clearing an upvote threshold, biasing the system in favor of the challengee and discouraging trolling behaviors.}
        \item Earning insufficient chore points in a month \textit{(passive)}
        \item Bonus hearts fade over time, back to baseline \textit{(passive)}
    \end{itemize}
\end{itemize}

While the semantics of chore points are simple and primarily economic, the semantics of hearts are nuanced and socially embedded. In general, the semantics of any token are a function of both the interactions possible with it, and the social perceptions of the community using it. In the case of \textit{Hearts}, the inclusion of multiple independent processes for gaining and losing hearts inhibits one-dimensional social interpretations, and causes hearts to be perceived closely to how they were intended: as a meaningful measure of contribution and conduct.

\vspace{2mm}

In particular, the regeneration of hearts over time has resulted in the striking phenomenon by which residents are able to experience the loss of hearts as \textit{meaningfully cathartic}, and the regeneration of hearts as \textit{meaningfully restorative}; at least in some cases, once hearts have regenerated, the transgression is experienced as forgiven. Residents have expressed that, with \textit{Hearts}, they feel less need to ``hold on'' to past ruptures. To borrow a concept from computer science, hearts become a ``psychological sink'' for intra-group tension, facilitating emotional sublimation and release. \citex{girard:1972} discusses the essential role of sublimating cultural processes in safely releasing intra-group tension: that this function could be performed at least in part by a cybernetic system should be seen as promising.

Recalling our frame of high- and low-order cognition, we suggest that the dynamic and regenerative nature of \textit{Hearts} performs \textit{computationally} the higher-order labor of remembering historical grievances, freeing residents to focus their energy on the lower-order work of identifying and proactively addressing norm violations as they occur, or of simply enjoying each other’s company.

\subsection{Things}
\vspace{2mm}

\textit{Things} is the spending and procurement mechanism of Chore Wheel. Simpler than \textit{Chores} and \textit{Hearts}, \textit{Things} demonstrates how the design language given in section \ref{principles} can be flexibly applied to a wide range of problems: in this case, helping residents collaboratively spend shared funds. The setup is simple: residents are given a monthly budget and can propose various types of purchases, with adaptive approval thresholds increasing with price.

The goal of \textit{Things} is to eliminate the need for any one individual to manage inventory (a high-order leadership task) by enabling individual residents to request supplies at the moment they perceive something running low (a low-order task leveraging local information). All residents can see current balances and can decide whether or not to approve a purchase, making spot judgments about the merit of the purchase vis-à-vis the amount of available funds. The adaptive threshold allows small purchases to be made with little to no engagement, but ensures broad-based support for larger or extraordinary purchases. As in \textit{Chores}, residents can propose amendments to the list of items, a temporary performance of leadership on behalf of the group.

\vspace{2mm}

In Spring 2024, named accounts were introduced to add an additional semantic layer to the purchase decision. Prior, there was only a single unnamed account. Residents would verbally express a desire to save for larger purchases, but the distributed decision-making would invariably result in the entire balance being spent by the end of each month. Introducing named accounts changed the semantics of purchasing decisions, as purchase proposals could be evaluated in a richer context, e.g. ``Is this a justified \textit{major} purchase?'' rather than the simpler ``Is this a justified purchase?'' This addressed the underlying issue not by increasing the difficulty of the decision problem, but by adding relevant local information -- an intervention consistent with our design principles.-blur
\section{Exploratory Data Analysis}
\label{analysis}

In this section, we present an exploratory analysis of data produced by Chore Wheel. The goal is not to make causal arguments, but to demonstrate the kinds of analyses made possible by this new source of data, and to support the narrative presented by this paper.

Chore Wheel is open-source and privacy-preserving, storing no personally-identifiable information.\footnote{The implementation is available at \texttt{github.com/zaratanDotWorld/choreWheel}} All data are anonymous.

\vspace{2mm}

\begin{figure}[ht]
    \centering
    \begin{subfigure}[b]{0.475\linewidth}
        \centering
        \includegraphics[width=\textwidth]{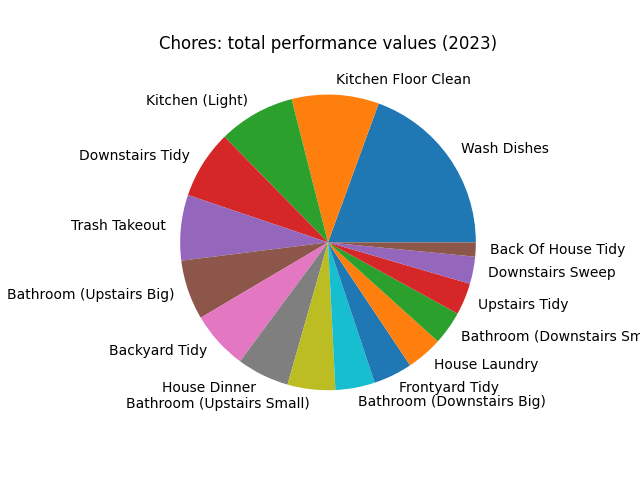}
        \caption{}
        \label{fig:chore-performances}
    \end{subfigure}
    \hfill
    \begin{subfigure}[b]{0.475\linewidth}
        \centering
        \includegraphics[width=\textwidth]{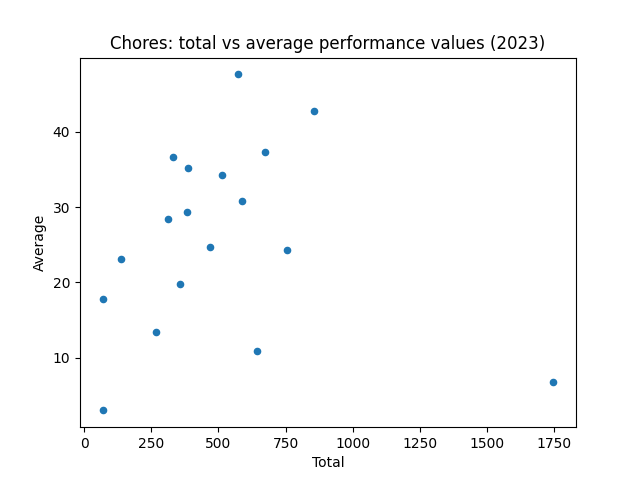}
        \caption{}
        \label{fig:chore-performances-2}
    \end{subfigure}
    \caption{Total performances by chore type for 2023 (n=567). Figure \ref{fig:chore-performances} shows kitchen-related tasks absorbing roughly 1/3 of the total points. Figure \ref{fig:chore-performances-2} shows ``Wash Dishes'' is a unique chore, absorbing the plurality of points despite a low average performance value.}
    \label{fig:total-performances}
\end{figure}

\textbf{Figure \ref{fig:total-performances}} shows how \textit{Chores} was used to organize regenerative labor. Figure \ref{fig:chore-performances} shows the proportions of \textit{all points} earned in the period, grouped by the chore they went to fund. We see that approximately 1/3 of all points were absorbed by the kitchen-related tasks of \textit{Wash Dishes}, \textit{Kitchen Floor Clean}, and \textit{Kitchen (Light)}, indicating that these tasks were seen as the highest-priority by the residents.

Figure \ref{fig:chore-performances-2} adds nuance, showing how the total performance values relate to the \textit{average} performance values of the same tasks. For most chores, a higher priority reflects a ``harder'' task which should be worth more points per-performance. However, in the case of the high-priority \textit{Wash Dishes}, the average per-performance value was low, indicating a simple task repeated frequently. These data correspond with our intuitions: while the kitchen floor should be cleaned perhaps twice per month, the dishes are typically done at least once per day.

\vspace{2mm}

\begin{figure}[ht]
\centering\includegraphics[width=1\linewidth]{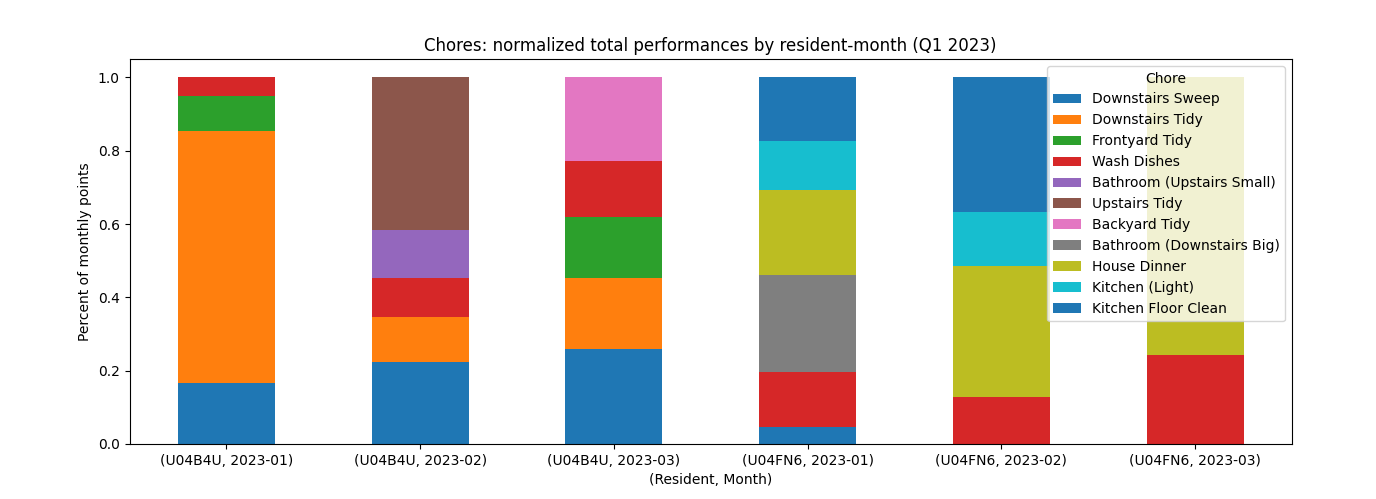}
\caption{Normalized monthly chore performances for two residents over three months. We see how, without explicit organization, resident U04B4U (left) self-sorts into common-area tasks, while U04FN6 (right) self-sorts into kitchen tasks. Normalization accounts for differences in total points per month caused by time out-of-town.}
\label{fig:resident-performances}
\end{figure}

\textbf{Figure \ref{fig:resident-performances}} reveals the ways residents self-specialize in particular tasks. One of the goals of \textit{Chores} was to give residents freedom and flexibility in meeting their obligations; an early open question wondered what dynamics might emerge. Here, we see clear personal preferences for certain tasks over others. In the period considered, resident U04B4U (chart left) earned the majority of their points through various sweeping and tidying tasks. Resident U04FN6 (chart right), on the other hand, earned the majority of their points by cleaning the kitchen and making house dinners. Both residents take at least a few turns doing the dishes. We can speculate that this cooperative specialization has led to greater efficiency in task performance, resulting in less time spent doing chores overall, but we lack the data to decide conclusively.

\vspace{2mm}

\begin{figure}[ht]
\centering\includegraphics[width=1\linewidth]{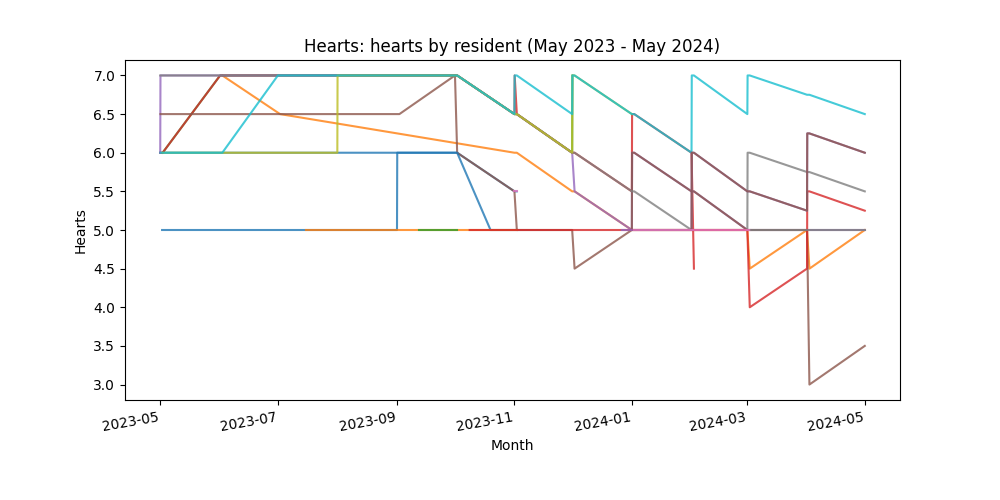}
\caption{Cumulative Hearts values for residents over 12 months (n=255). We see hearts reflecting both high-engagement behaviors resulting in bonus hearts, as well as low-engagement behaviors resulting in loss of hearts. The fading of bonus hearts was introduced in October 2023, and the rate of fading was reduced in April 2024.}
\label{fig:hearts-values}
\end{figure}

\textbf{Figure \ref{fig:hearts-values}} gives a lens into resident's norm-compliance over time, as seen through \textit{Hearts}. Over the period considered, we see residents earning karma, shirking chores, and experiencing returns to ``social baseline'' over time. Notable in these data are the ways that high-engaging residents who regularly earn karma experience frequent cycles of losing hearts (due to fading karma) only to re-earn  them in the following month. In contrast, low-engaging residents who lose hearts due to chore shirking seem to rarely repeat those behaviors. This suggests that earning hearts through karma is experienced as a meaningful recognition, and that losing hearts due to chore shirking is experienced as a meaningful sanction.

\vspace{2mm}

\begin{figure}[ht]
    \centering
    \begin{subfigure}[b]{0.475\linewidth}
        \centering
        \includegraphics[width=\textwidth]{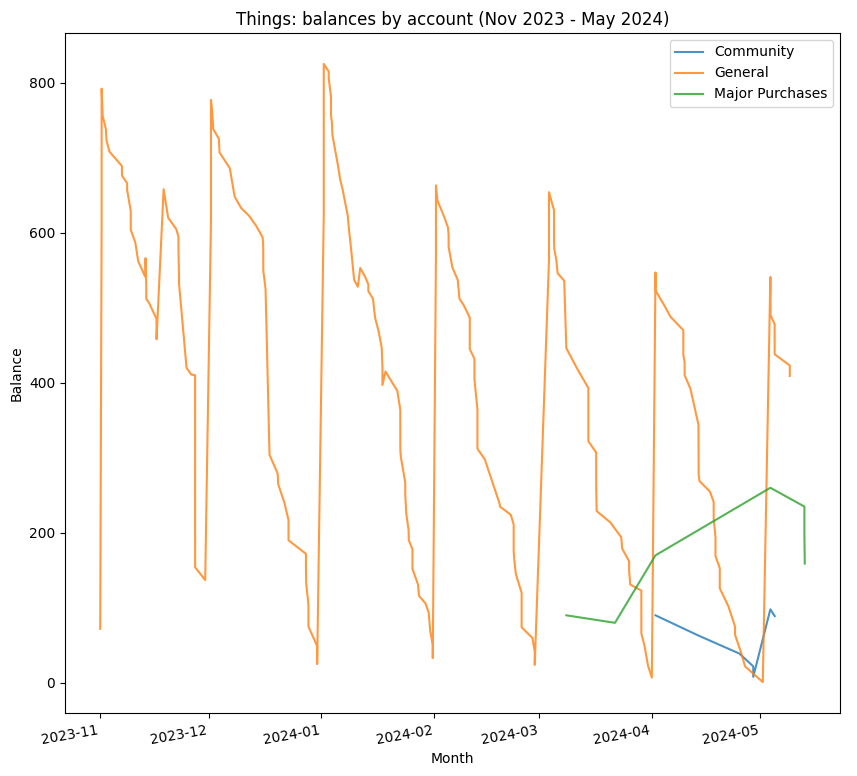}
        \caption{}
        \label{fig:things-balances}
    \end{subfigure}
    \hfill
    \begin{subfigure}[b]{0.475\linewidth}
        \centering
        \includegraphics[width=\textwidth]{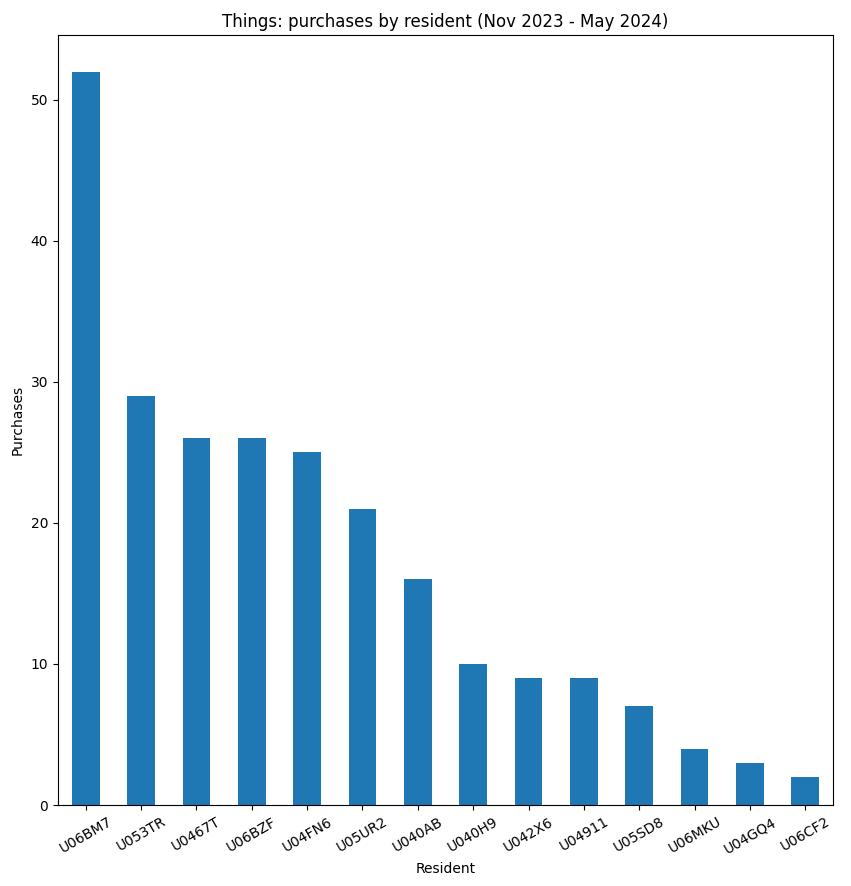}
        \caption{}
        \label{fig:things-buyers}
    \end{subfigure}
    \caption{\textit{Things} purchasing behaviors (n=551). Figure \ref{fig:things-balances} shows account balances over 6 months, revealing savings behavior in some accounts and not others. Named accounts were introduced in Spring 2024. Figure \ref{fig:things-buyers} shows per-resident purchases during the same period, revealing a heavy-tailed power-law distribution.}
    \label{fig:things-plots}
\end{figure}

\textbf{Figure \ref{fig:things-plots}} details \textit{Things} purchasing behaviors. Figure \ref{fig:things-balances} shows how different accounts are used by residents. We see how the General account is consistently spent down month-over-month, while the Major Purchases account, which is mechanistically identical, exhibits savings behavior.

Figure \ref{fig:things-buyers} shows the distribution of purchasing activity per-resident. We see a power-law distribution with a heavy tail: 80\% of all purchases are performed by the top 42\% of residents, more than double the rate of participation predicted by the typical 80/20 power-law distribution (\citey{newman:2005}). This relatively egalitarian engagement pattern may reflect greater feelings of personal agency among residents, although more data are needed to decide conclusively.
\section{Discussion}
\label{discussion}

While the example of Sage provides useful signal, open questions remain.

\vspace{2mm}

\textbf{The first} is the relative contribution to Sage's stability of the computational system compared to the particular people that live there. While Chore Wheel is unarguably useful, Sage has also housed very capable individuals. What is the role of abstract structure, and what is the role of specific culture, in explaining Sage's performance?

Drawing strong conclusions from non-experimental field data is difficult. Analyzing data from a corpus of 5,000 independently-operated Minecraft servers, \citex{frey:2019} argue that while it is impossible to \textit{prove} that complex governance schemes are needed to support large player bases, it is difficult to imagine the causality being reversed. As the authors suggest, the counter-argument that large player bases \textit{spontaneously generate} complex governance schemes does not hold water.

Our study of Sage House is based on 20 months of field data, with many possible confounds. That said, we feel justified in concluding that Sage’s continuous operation, maintaining high occupancy and low overhead while serving a population broadly inexperienced in communal living, represents a validation of Chore Wheel’s theory and practice.

\vspace{2mm}

\textbf{The second} open question considers HCI, or human-computer interaction. The pairwise allocation mechanism described in \citex{kronovet:2018} was well-received on publication, particularly within the Web3 / Public Goods Funding communities. In the years since, two independent engineering organizations, General Magic and dOrg,\footnote{Pairwise (\texttt{pairwise.vote}) and Pairdrop (\texttt{pairdrop.daodrops.io}), respectively.} have developed front-end interfaces for the mechanism, indicating that the design space is large.

Chore Wheel’s implementation, in the form of a Slack app, is not yet fully intuitive. Residents report that while they are able to understand the chore prioritization mechanism after a brief in-person explanation, the initial unsupported encounter was often confusing. Ultimately, the potential of this mechanic can only be realized to the extent that the interface is well-designed. Exploring the design space of human-computer interfaces, in particular as they relate to the pairwise allocation mechanism, remains an important line of research.

\vspace{2mm}

\textbf{The third}  open question relates to the value of ``struggle,'' or of the \textit{training value} of performing collective / higher-order cognition. If human relationships are \textit{too} mediated by computational tools, the argument goes, individual capacity may decline. Discussing Web3 technological maximalism, \citex{lotti:2024} characterize this as ``Crypto’s Three-Body Problem.'' Building on \citex{lessig:1998}, the authors argue that, by mediating relationships with code, the Web3 community has left itself \textit{structurally unable} to overcome its social coordination challenges: without the ability to produce and maintain social norms, computers have made things more fragile, not less.

\vspace{2mm}

As discussed in section \ref{mechanism}, distributed decision-making in \textit{Things} made it difficult for residents to coordinate around large purchases. While individuals expressed wanting to save for larger purchases, supplemental institutions such as shared spreadsheets for long-term planning were never produced. Only after support for named accounts was added to the system via constitutional action did savings behavior emerge.

It is also important to acknowledge that, while this paper has emphasized Sage's computational infrastructure, residents \textit{do} engage in supplemental in-person process. Each month residents gather for an informal meeting, creating opportunities to flexibly and synchronously discuss residual issues not comprehended by the Chore Wheel system. This time, often used to re-affirm shared values and pro-actively address emerging problems, plays what is likely an essential complementary role to the house's technical apparatus.

\vspace{2mm}

It is exposure to stressors which make organisms stronger. In the end, we may be forced to reckon with a type of cybernetic ``strange loop,'' through which more effective regulatory systems prevent participants from developing the capacity to productively engage with the systems in the first place. As \citex{scott:1998} writes, all formal systems are dependent on informal cultural capacities which the formal system itself cannot create nor maintain. The in-person meeting may be performing important secondary social functions, which we unwittingly lose when deciding through screens. The pursuit of institutional resilience may be self-limiting.

On the other hand, popular elections and parliamentary procedures can be seen as types of computer programs, implemented with pen-and-paper instead of silicon. Pass-fail voting is not the end of intellectual history, and ``technology'' broadly defined has been essential in helping structure and sustain complex society. Finding the ``right balance'' of human and machine is a hard problem and an important line of research.

\vspace{2mm}

\textbf{Returning to our motivating themes} of leadership, organizational resilience, and cybernetics, we conclude that while limited and incomplete, these results are promising. While the role of leadership has not been eliminated and likely will never be, we have shown that tasks which have historically required personal leadership can be accomplished, with some constraints, by non-leaders enabled by appropriate technical systems. The case study of Sage House has shown that, in practice, such human-computer systems can function as well or better than pure-human alternatives. Overall, we believe that this inquiry into \textit{distributed digital institutions} represents a valuable exploration which we should continue to pursue.




\vspace{2mm}
\vspace{2mm}

\footnotesize{
    \noindent \textbf{Ethics Statement}: All data from Chore Wheel are stored anonymously. Explicit consent was received when referencing conversations with housemates. Sage House is co-owned by Daniel and Robert Kronovet. Chore Wheel is primarily developed by Kronosapiens Labs, owned by Daniel, with support from Gitcoin Grants rounds 18, 19, and 20.
}

\end{document}